\documentclass[superscriptaddress, preprintnumbers,amsmath,amssymb,twocolumn]{revtex4}
\usepackage{graphicx}
\usepackage{dcolumn}
\usepackage{bm}
\usepackage{latexsym}
\usepackage{float}
\begin{document}

\title{Collective cargo hauling by a bundle of parallel microtubules: bi-directional motion caused by load-dependent polymerization and depolymerization }

\author{Dipanwita Ghanti}
\affiliation{Department of Physics, Indian Institute of Technology Kanpur, 208016}
\author{Debashish Chowdhury{\footnote{Email: debch@iitk.ac.in}}}
\affiliation{Department of Physics, Indian Institute of Technology Kanpur, 208016}
\begin{abstract}
A microtubule (MT) is a hollow tube of approximately 25 nm diameter. The two ends of 
the tube are dissimilar and are designated as `plus' and `minus' ends. Motivated by the 
collective push and pull exerted by a bundle of MTs during chromosome segregation in 
a living cell, we have developed here a much simplified theoretical model of a bundle of 
parallel dynamic MTs. The plus-end of all the MTs in the bundle are permanently attached 
to a movable `wall' by a device whose detailed structure is not treated explicitly in our 
model. The only requirement is that the device allows polymerization and depolymerization 
of each MT at the plus-end. In spite of the absence of external force and direct 
lateral interactions between the MTs, the group of polymerizing MTs attached to the 
wall create a load force against the group of depolymerizing MTs and vice-versa; the load 
against a group is shared equally by the members of that group. Such indirect interactions 
among the MTs gives rise to the rich variety of possible states of collective dynamics that we 
have identified by computer simulations of the model in different parameter regimes. The 
bi-directional motion of the cargo, caused by the load-dependence of the polymerization 
kinetics, is a ``proof-of-principle'' that the bi-directional motion of chromosomes before 
cell division does not necessarily need active participation of motor proteins.

\end{abstract}

\maketitle

\section{Introduction}

A microtubules (MT) is a hollow tube of approximately 25 nm diameter. 
It is one of major components of the cytoskeleton \cite{howard01} 
that provides mechanical strength to the cell. Each MT is polar in the 
sense that its two ends are structurally as well as kinetically dissimilar. 
One of the unique features of a MT in the intracellular environment 
is its dynamic instability \cite{desai97}. 
The steady growth of a polymerizing MT takes place till a ``catastrophe'' 
triggers its rapid depolymerization from its tip. Often the depolymerizing 
MT is ``rescued'' from this decaying state before it disappears completely 
and keeps growing again by polymerization till its next catastrophe. Thus, 
during its lifetime, a MT alternates between the states of polymerization 
(growth) and depolymerization (decay).
  
One of the key structural features of a MT is that during its depolymerization 
the tip of this nano-tube is curved radially outward from its central axis.  
While polymerizing, the growing MT can exert a pushing force against  
a transverse barrier thereby operating, effectively, as a nano-piston 
\cite{oster04,laan08}. 
Although lateral cross linking between the MTs and their unzipping can 
have interesting effects \cite{kuhne09,krawczyk11}, no such cross link 
is incorporated in our model here because of the different motivation of 
our work.
Similarly, the splaying tip of a depolymerizing MT can pull an object in a 
manner that resembles the operation of a nano-hook 
\cite{peskin95,mcintosh10}.
Thus, a MT can perform mechanical work by transducing input chemical 
energy. In analogy with motor proteins that transduce chemical input 
energy into mechanical work, force generating polymerizing and 
depolymerizing MTs are also referred to as molecular motors 
\cite{chowdhury13a,chowdhury13b,kolomeisky13}.

In this paper we consider a {\it bundle} of parallel MTs that are {\it not} 
laterally bonded to each other. Such bundles have been the focus of attention 
in recent years because of the nontrivial collective kinetics and 
force generation by a bundle in spite of the absence of any direct 
lateral bond between them \cite{laan08}.
For example, while driving chromosomal movements in a mammalian cell before 
cell division, the members of a bundle that is attached to a single 
kinetochore wall, undergo catastrophe and rescue in a synchronized manner 
\cite{mcintosh12}. 
The type of collective phenomenon under consideration here is of current 
interest also in many related or similar contexts in living systems. 
For example, there are similarities between the motion of a chromosomal 
cargo pushed/pulled by a group of polymerizing/depolymerizing MTs 
and that of a membrane-bounded vesicular cargo hauled by the members of two 
antagonistic superfamilies of molecular motors along a filamentous track 
\cite{gross04,welte04,berger11}. But, as we explain here, there are crucial 
differences between the two systems. Other examples of similar collective 
dynamics include (stochastic) oscillations in muscles, flagella (a cell appendage) 
of sperm cells, etc. \cite{chowdhury13a,chowdhury13b}.

Here we develop a theoretical model for the collective kinetics of a 
{\it bundle} of parallel MTs that interact with a movable wall. The model is based on 
minimum number of hypotheses for the polymerization kinetics of individual MTs. 
More specifically, the model takes into account the force-dependence of the  
rates of catastrophe/rescue and polymerization/depolymerization. These hypotheses 
are consistent with {\it in-vitro} experiments on single MT reported so far in 
the literature. Numerical simulation of the model demonstrates synchronized 
growth and shrinkage of the MT bundle in a regime of model parameters. 
We classify the states of motion of the 
wall that is collectively pushed and pulled by the MT bundle. In one of the states 
thus discovered the wall executes {\it bi-directional} motion and it arises from 
a delicate interplay of load-force-dependent rates of MT polymerization and 
depolymerization. This work also provides direct evidence that bidirectional 
motion of MT-driven cargo is possible even in the absence of any motor protein 
\cite{chowdhury13a,chowdhury13b}; the ``proof-of-principle'' presented here will be elaborated later 
elsewhere \cite{ghanti14b} in the context of  modeling chromosome segregation.

\section{The model}

\begin{figure}[H]
\includegraphics[angle=90,width=0.85\columnwidth]{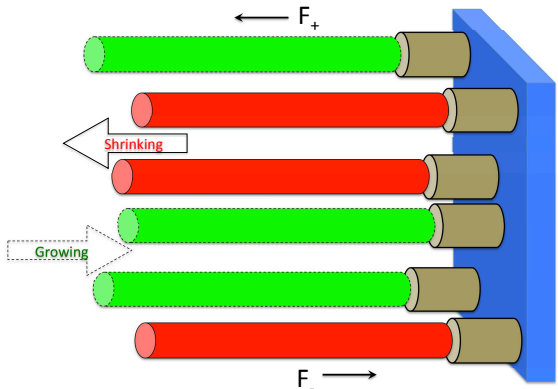}\\
\caption{(Color online) Our model is depicted schematically. 
The wall to which the MTs are attached is represented by the 
blue vertical slab. The polymerizing (growing) and depolymerizing 
(shrinking) MTs are represented by the horizontal green cylinders 
(with dashed outlines) and red cylinders (with solid outlines), 
respectively. The grey short cylinders, into 
which the plus end of the MTs are inserted, represent the 
``couplers'' that couple the MTs with the wall. The directions of 
polymerization and depolymerization are indicated by the 
double arrows. The directions of the load force (the absolute 
value of which is $F_{+}$) on the polymerizing MTs and load 
tension (the absolute value of which is $F_{-}$) on the 
depolymerizing MTs are indicated by the single arrows.  }
\label{fig-multMTmodel}
\end{figure}

As we stated in the introduction, our study has been motivated partly by 
collective force generation by MTs during chromosome segregation. 
The model is depicted schematically in fig.\ref{fig-multMTmodel}.
At present the identity of the molecules that couple MTs to the chromosomal 
cargo and the detailed mechanism of force generation at the MT-cargo 
interface are under intense investigation   
\cite{mcintosh12,cheeseman08}. 
Plausible scenarios for the coupling suggested in the literature include, 
for example, mechanisms based on (i) biased diffusion of the coupler 
\cite{hill85}, (ii) power stroke on the coupler \cite{koshland88}, 
(iii) attachment (and detachment) of long flexible tethers 
\cite{zaytsev13}.
Moreover, the mechanisms of the MT-cargo coupling may also vary from one 
species to another. 
Therefore, in order to maintain the generic character of our model, we neither  
make any explicit microscopic model for the coupler nor postulate any detailed 
mechanism by which the individual members of the MT bundle couple with the 
chromosomal cargo. This is similar, in spirit, to the kinetic models of 
hauling of membrane-bounded cargo where no explicit molecular mechanism is 
assumed for the motor-cargo coupling \cite{berger11}.
The only assumption we make about the MT-wall coupling is that none of the 
MTs detaches from the wall during the period of our observation; catastrophe 
and rescue of a MT merely reverse the direction of the force that it exerts on 
the wall. 

We assume that during the period of our observation neither nucleation 
of new MTs take place nor does any of the existing MTs disappear altogether 
because of its depolymerization. Therefore, the total number of MTs, 
denoted by $N$, is conserved. We use the symbols $n_{+}(t)$ and $n_{-}(t)$ 
to denote the MTs in the growing and shrinking phases, respectively, at 
time $t$ so that $n_{+}(t)+n_{-}(t)=N$. Thus, at time $t$, $n_{+}(t)$ MTs push 
the cargo while, simultaneously, $n_{-}(t)$ MTs pull the same cargo in the 
opposite direction. 

As we show in this paper, the distribution of the MTs in the two groups, 
namely $n_{+}$ and $n_{-}$, is a crucial determinant of the nature of the 
movements of the wall. By carrying out computer simulations, we also 
monitor the displacement and velocity distribution of the wall in different 
regimes of the model parameters to characterize its distinct states of 
motility. The observed states of motility are displayed on a phase diagram.

\subsection{Parameters and equations: force-dependence} 

The force-dependent rates of catastrophe and rescue of a MT are denoted by 
the symbols $c(F)$ and $r(F)$, respectively, while $c_{0}$ and $r_{0}$ 
refer to the corresponding values in the absence of load force. 
The experimental data collected over the last few years \cite{franck07} 
and very recent modeling \cite{sharma14} have established the dependence 
of the rates of catastrophe and rescue on the load force. Accordingly, 
we assume 
\begin{eqnarray}
c(F) = c_0 ~exp(|F|/F_{*c})  
\label{eq-ppmFa}
\end{eqnarray}
where $|F|$ is the absolute value of the load force opposing polymerization, 
and $F_{*c}$ is the characteristic force that characterizes the rapidity of 
variation of the catastrophe rate with the load force. Similarly, we assume 
\begin{eqnarray}
r(F) = r_0 ~exp(|F|/F_{*r})  
\label{eq-ppmFb}
\end{eqnarray}
where $|F|$ is the absolute value of the load tension opposing 
depolymerization and $F_{*r} $ is the characteristic force that characterizes 
the rapidity of variation of the depolymerization rate with the load force.

We denote the load-free velocities of polymerization and depolymerization of a  
single MT by the symbols $v_{g}$ and $v_{d}$, respectively. For the load-dependence 
of these velocities we follow the standard practice in the literature for modeling 
the load-dependence of the velocity of motor proteins.  Specifically, we assume linear 
force-velocity relation 
\begin{equation}
   v_{+}= v_{g}\biggl(1-\frac{|F_{+}|}{F_{s_{+}}}\biggr) ~ {\rm for}~ |F_{+}| \leq F_{s_{+}}
\label{eq-vplusF}
  \end{equation}
for the polymerizing MTs, and 
\begin{equation}
  v_{-}= v_{d}\biggl(1-\frac{|F_{-}|}{F_{s_{-}}}\biggr)  ~ {\rm for}~ |F_{-}| \leq F_{s_{-}}
\label{eq-vminusF}
\end{equation} 
for the depolymerizing MTs, where $|F_{+}|$ and $|F_{-}|$ denote the 
absolute values of the respective load forces opposing polymerization 
and depolymerization, respectively, of the corresponding MT. 
$F_{s_{+}}$ and $F_{s_{-}}$ are the absolute values of the {\it stall forces}
corresponding to the polymerizing and depolymerizing MTs.

In the spirit of many earlier theoretical models of bidirectional transport of 
membrane-bounded cargo \cite{berger11} 
we also make the following assumptions:\\ 
(i) push on the wall by polymerizing MTs generates the indirect load tension on 
the depolymerizing MTs while the pull of the latter on the same wall 
simultaneously creates the load force against the polymerizing MTs. 
(ii) the load force against polymerization is shared equally by the $n_{+}(t)$ 
MTs, and the load tension against depolymerization is also shared equally 
by the $n_{-}(t)$ MTs.
At any arbitrary instant of time $t$, each of the $n_{+}(t)$ polymerizing MTs 
attached the chromosomal cargo experiences a load force $F_{+}$ and exerts 
a force $-F_{+}$. Similarly, each of $n_{-}(t)$ depolymerizing MTs feels a load 
tension $-F_{-}$ and offers a load force $F_{-}$. Therefore, force balance 
on the wall attached simultaneously by $ n_{+}(t)$ and $n_{-}(t)$ MT is
\begin{equation}
n_{+}F_{+}=-n_{-}F_{-}\equiv F_{w}(n_{+},n_{-}).
\label{eq-npnmF}
\end{equation}
Equation (\ref{eq-npnmF}), which defines the symbol $F_{w}(n_{+},n_{-})$, 
is just a mathematical representation of Newton's third law: the force 
exerted by the wall on each MT is equal and opposite to that exerted by 
the same MT on the wall. According to our choice of the signs, the load 
force on the polymerizing MTs are positive. 
Based on the assumption (ii) above, we now have 
\begin{eqnarray} 
|F_{+}(n_{+},n_{-})|&=&|F_{w}(n_{+},n_{-})|/n_{+} \nonumber \\
|F_{-}(n_{+},n_{-})|&=&|F_{w}(n_{+},n_{-})|/n_{-}.  
\label{eq-Fpm}
\end{eqnarray}
When $n_{+}(t)$ and $n_{-}(t)$ are the numbers of MTs in the polymerizing and 
depolymerizing phases, respectively, the corresponding catastrophe and rescue rates are  
\begin{eqnarray} 
 c_{n+,n-}(F) &=& c_{0}~ exp(|F_{+}(n_{+},n_{-})|/F_{\star c}) \nonumber \\
 r_{n+,n-}(F) &=& r_{0}~ exp(|F_{-}(n_{+},n_{-})|/F_{\star r}), 
\label{eq-ppmFfinal}
\end{eqnarray}
respectively.

Since all the $n_{+}(t)$ polymerizing MTs and $n_{-}(t)$ depolymerizing MTs are, 
by definition, attached to the wall at time $t$, their velocities of 
growth and decay, respectively, must be identical to the wall velocity 
$v_{w}$, i.e., 
\begin{equation}
v_{w}(n_{+}(t),n_{-}(t)) = v_{+}(F_{w}/n_{+}(t)) = - v_{-}(-F_{w}/n_{-}(t))
\label{eq-vequality}
\end{equation} 
Now substituting eqs.(\ref{eq-Fpm}) into eqs.(\ref{eq-vplusF}) and (\ref{eq-vminusF}) we get
\begin{equation}
  v_{+}= v_{g}\biggl(1-\frac{F_{w}}{n_{+}F_{s_{+}}}\biggr) 
\label{eq-vplus}
\end{equation} 
\begin{equation}
  v_{-}= v_{d}\biggl(1-\frac{F_{w}}{n_{-}F_{s_{-}}}\biggr)
\label{eq-vminus}
\end{equation}

Imposing the constraint (\ref{eq-vequality}) on eqs. (\ref{eq-vplusF}) and (\ref{eq-vminusF})  we get 
the force 
\begin{equation}
   F_{w}(n_{+},n_{-})= \mu n_{+}F_{s_{+}}+(1-\mu)n_{-}F_{s_{-}}
\label{eq-Fw}
\end{equation}
and velocity 
\begin{equation}
   v_{w}(n_{+},n_{-})= \frac{n_{+}F_{s_{+}}-n_{-}F_{s_{-}}}{n_{+}\biggl(\frac{F_{s_{+}}}{v_{g}}\biggr)+n_{-}\biggl(\frac{F_{s_{-}}}{v_{d}}\biggr)}
\label{eq-vw}
\end{equation}
for the wall where $ \mu^{-1}=1+\frac{(n_{+}F_{s_{+}}v_{d})}{(n_{-}F_{s_{-}}v_{g})} $.

At every instant of time $t$ the state of the system is characterized by $ n_{+}(t)$ and $n_{-}(t)$. 
The probability of finding the system in the state $n_{+}(t),n_{-}(t)$ at time $t$ is denoted by 
$P(n_{+},n_{-},t)$ and its time evolution is governed by the master equation 
\begin{eqnarray}
 && \frac{dP(n_{+},n_{-},t)}{dt} = \nonumber \\  
&+& c_{n_{+}+1,n_{-}-1}~ P(n_{+}+1,n_{-}-1,t) \nonumber \\ 
&+& r_{n_{+}-1,n_{-}+1}~ P(n_{+}-1,n_{-}+1,t) \nonumber \\ 
&-& (c_{n_{+},n_{-}}+r_{n_{+},n_{-}})   P(n_{+},n_{-},t) \nonumber \\
\label{eq-master1}
 \end{eqnarray}
where $c_{n_{+},n_{-}}$ and $r_{n_{+},n_{-}}$ are given by the equations 
(\ref{eq-ppmFfinal}) while 
\begin{eqnarray} 
 c_{n_{+}+1,n_{-}-1}(F) &=& c_{0}~ exp(|F_{+}(n_{+}+1,n_{-}-1)|/F_{\star c}) \nonumber \\
 r_{n_{+}-1,n_{-}+1}(F) &=& r_{0}~ exp(|F_{-}(n_{+}-1,n_{-}+1)|/F_{\star r}). \nonumber \\ 
\label{eq-ppmFfinal2}
\end{eqnarray}
Note that the rate constants in equation (\ref{eq-master1}) depend on the 
time-dependent quantities $n_{+}(t)$ and $n_{-}(t)$. Therefore, the rate 
constants on the right hand side of eqnation (\ref{eq-master1}) take their 
instantaneous values at time $t$.

\subsection{Comparison with other similar phenomena and models} 

We can now compare the model system under study here with the models of 
transport of membrane-bounded cargo (vesicles or organelles) by 
antagonistic motor proteins (e.g., kinesins and dyneins) that move on a 
filamentous track (e.g., a MT) \cite{chowdhury13a,chowdhury13b}. 
The polymerizing and depolymerizing MTs are the analogs of kinesins and 
dyneins. However, one crucial difference between the two situations is that 
unlike a MT, that can switch from polymerizing to depolymerizing phase 
(because of catastrophe) and from depolymerizing to polymerizing phase 
(because of rescue), interconversion of plus and minus-end directed motor  
proteins is impossible. 

A systematic comparison of these two systems is presented in table \ref{table-comp}. 
\begin{widetext}

\begin{table}[H]
\begin{tabular}{|c|c|c|} \hline
Object or property & Multi-MT system & Multi-motor system \\\hline \hline
Cargo & kinetochore (kt) & vesicle or organelle \\\hline
Antagonistic force generators & Polymerizing MT, depolymerizing MT & kinesin, dynein \\\hline
Inter conversion of force generators &  Polymerizing MT ${\rightleftharpoons}$ Depolymerizing MT: possible & Kinesin ${\rightleftharpoons}$ Dynein: impossible \\\hline 
Total number conservation & $N_{+}(t)+N_{-}(t)=N$=constant & $N_{+}+N_{-}=N$=constant \\\hline
Numbers of + and - force generators  & Time-dependent $N_{+}(t)$, $N_{-}(t)$ & $N_{+}$=constant, $N_{-}$=constant \\\hline
Number attached to cargo & $n_{+}(t)= N_{+}(t), n_{-}(t)=N_{-}(t)$ & $n_{+}(t) \leq N_{+}, n_{-}(t) \leq N_{-}$ \\\hline 
Track & No analog & MT \\\hline
Rate of attachment to track & No analog & $\omega_{a}$ \\\hline
Rate of detachment from track & No analog & $\omega_{d}$ \\\hline
\end{tabular}
\caption{Comparison of cargo hauling by bundle of parallel dynamic MTs and by 
antagonistic cytoskeletal motor proteins.  
 }
\label{table-comp}
\end{table}

\end{widetext}

\section{Results and discussion: motility states}

We have simulated the time evolution of $n_{+}(t)$ and $n_{-}(t)$, that is described by 
the master equation (\ref{eq-master1}) using Gillespie algorithm \cite{gillespie07}. 
We get the instantaneous velocity $v_{w}(n_{+},n_{-})$ by substituting the corresponding 
values of $n_{+}(t)$ and $n_{-}(t)$ into equation (\ref{eq-vw}). This data is used not 
only for plotting the velocity distribution but also for computing the displacements of 
the wall as a function of time. Moreover, the instantaneous force $F_{w}(n_{+},n_{-})$ 
evaluated using (\ref{eq-Fw}) is used for updating $n_{+}$ and $n_{-}$ in the next time 
step. Finally, all the distinct states of motility of the wall are characterized by the statistics 
of $n_{+},n_{-}$ . For the simulation we have assumed that initially half of the MTs are in 
the the state of polymerization while the remaining half are in the state of depolymerization. 
For the presentation of our data, it would be convenient to reduce the number of 
parameters. Therefore, in this paper we report the results only for the cases where  
$v_{g} = v_{d}$,  $ c_{0} = r_{0} = \gamma_{0}$, $ F_{s+} = F_{s-} = F_{s}$. 
The magnitudes of $v_{g}=v_{d}=50$~nm/s are identical in all the 
figures plotted below.

\begin{figure}[H]
(A)\\
\includegraphics[angle=-90,width=0.85\columnwidth]{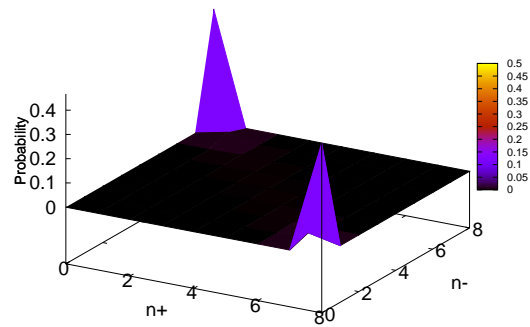} \\
(B)\\
\includegraphics[width=0.85\columnwidth]{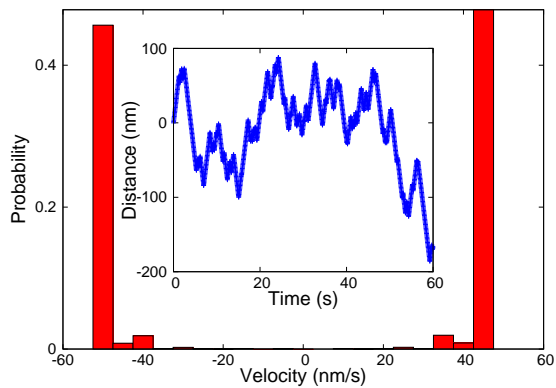}\\
\caption{Characteristics of ``bidirectional motion'' (+-) state. 
(A) There are two distinct maxima  in the 3D plot of the probability against the numbers of growing ($n_{+}$) and shrinking ($n_{-}$) MT; the peak near $n_{-}=0$ occurs when majority of the MTs are growing whereas the peak near $n_{+}=0$ indicates majority of the shrinking MTs. 
(B) The long stretches of  alternate positive and negative displacements are clearly visible in the typical trajectory displayed in the inset. 
The two distinct and sharp peaks at the velocities $\pm 50$nm/s correspond to the two long stretches of positive and negative displacements of the wall. Numerical value of the parameters, used in the simulation are $ F_{s}=6$~pN,$ v_{g}=v_{d}=$50 nm/s, $ c_{0} = r_{0} = 10$ s$^{-1}$, and $ F_{*c} = F_{*r} =3.0$~pN
}
\label{fig_m+-}
\end{figure}

\begin{figure}[H]
(A)\\
\includegraphics[angle=-90,width=0.85\columnwidth]{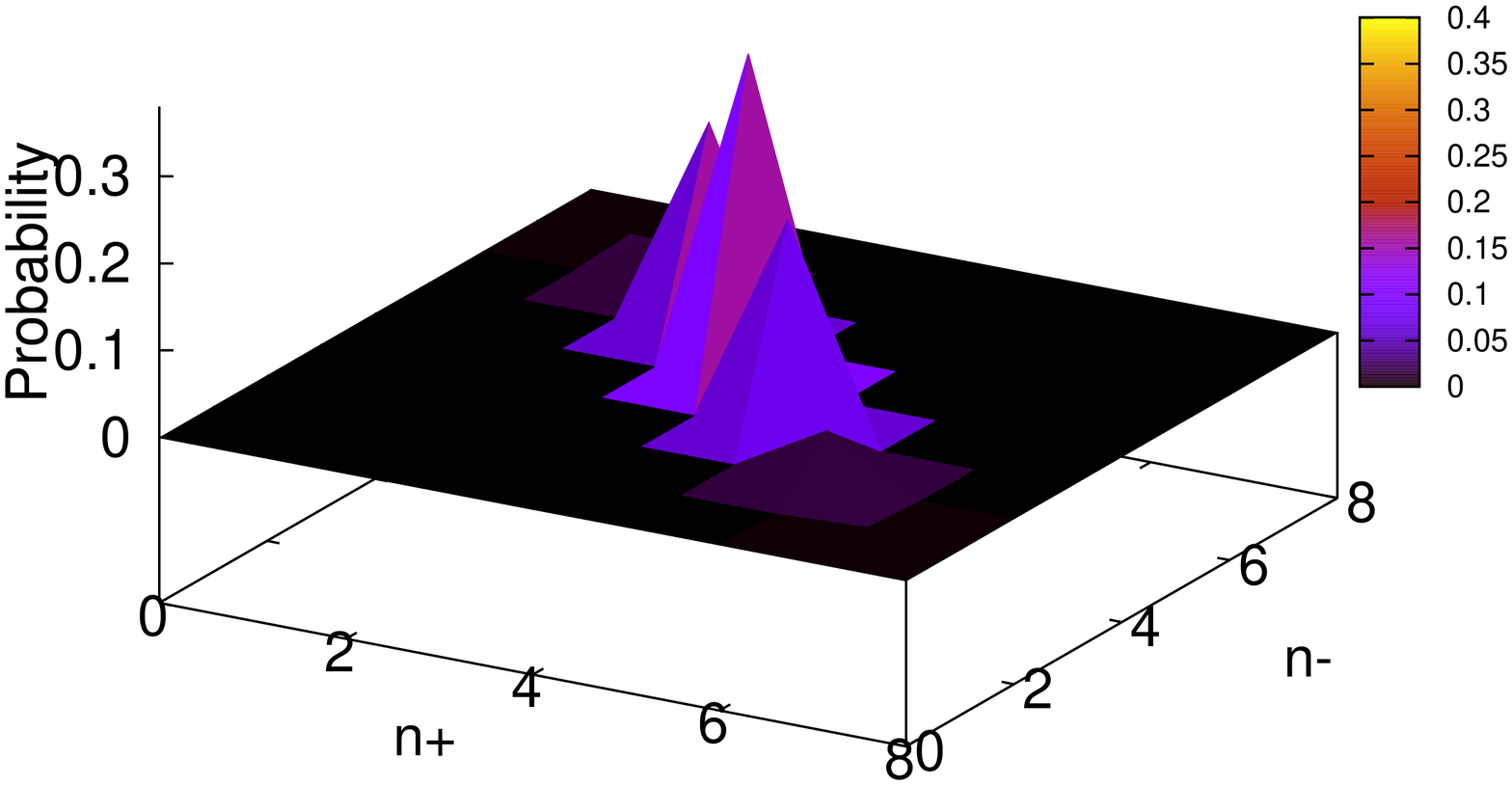}\\
(B)\\
\includegraphics[width=0.85\columnwidth]{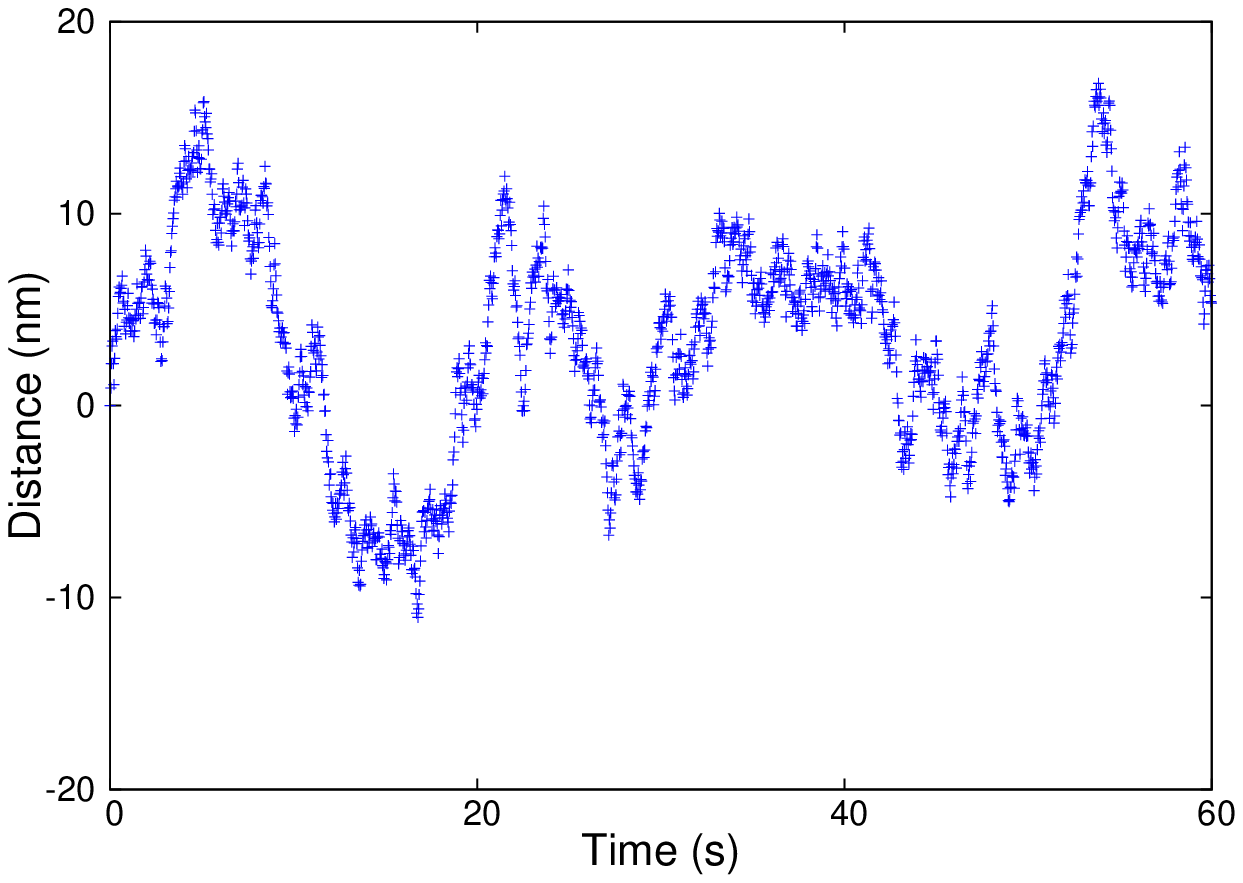}\\
(C)\\
\includegraphics[width=0.85\columnwidth]{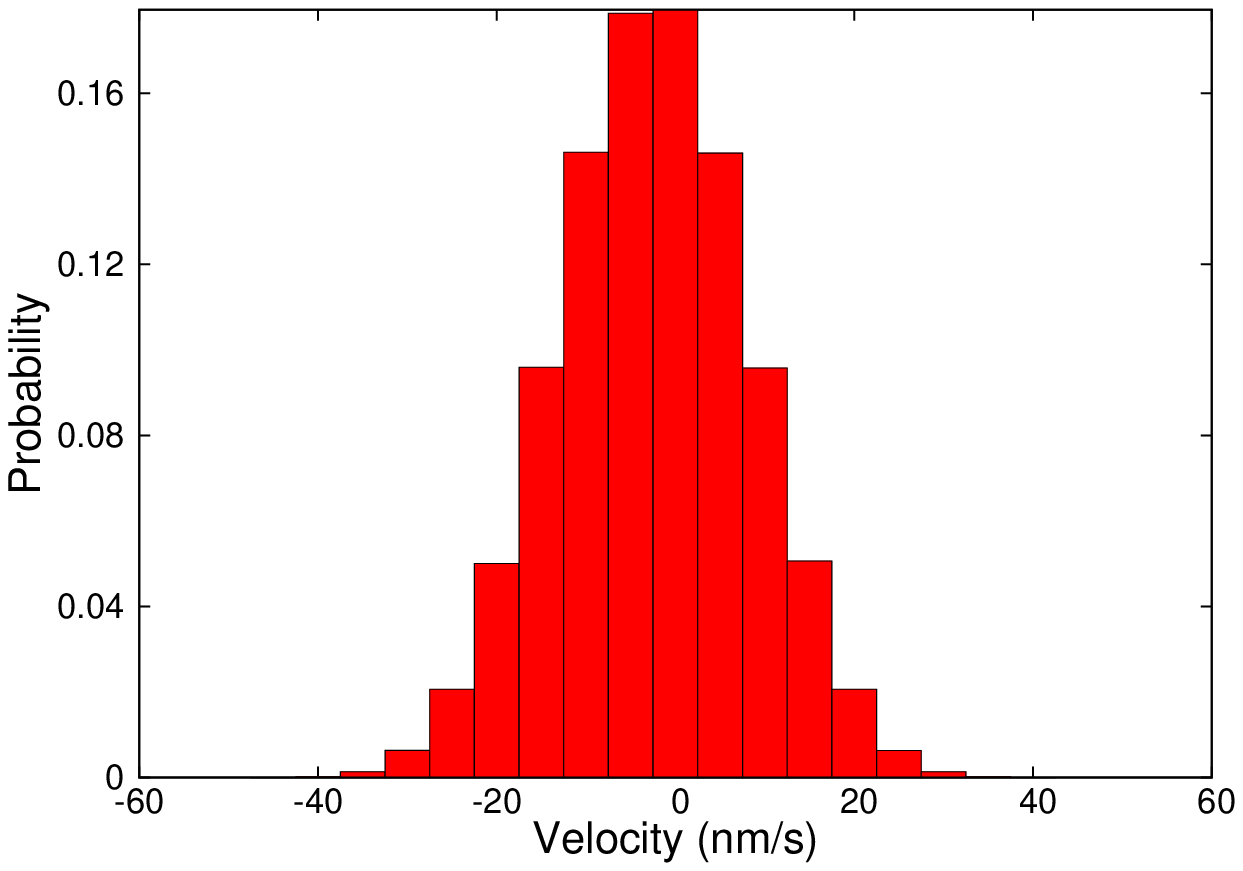}
\caption{Characteristics of ``no motion'' (0) state. 
(A) In the 3D plot of the probability against the numbers of growing ($n_{+}$) and shrinking ($n_{-}$) MTs peaks are seen along the diagonal, i.e., $n_{+}=n_{-}$. (B) The small random positive and negative displacements of the wall recorded with the passage of time reflects the fluctuations in the position of the wall as a consequence of the tug-of-war between the growing and shrinking MTs. (C) The sharp dominant peak at zero velocity in the distribution of velocities is a signature of the `no motion' state; the small non-zero probabilities at other velocities correspond to the small random excursions of the wall about its mean position. Numerical value of the parameters, used in the simulation are $ F_{s}=0.1$~pN,$ v_{g}=v_{d}=$50 nm/s, $ F_{*c} = F_{*r} =10.0$~pN and $ c_{0} = r_{0} = 2000 s^{-1}$. }
\label{fig_m0}
\end{figure}

$\bullet$ {\bf Bi-directional motion in the `Plus and Minus Motion (+-)' state}

First we simulated the model for the parameter values $ c_{0} = r_{0} = 10$ s$^{-1}$, and 
$ F_{*c} = F_{*r} =3.0$~pN. The numerical data obtained from simulation are plotted in 
fig.\ref{fig_m+-}. In this case the typical trajectories provide direct evidence for bi-directional 
motion of the wall. This is also consistent with the two distinct and sharp peaks at 
$V=\pm 50$ nm/s in the distribution of velocities of the wall. The physical origin of this 
bi-directional motion \cite{berger11} can be inferred from the nature of the distribution 
$P(n_{+},n_{-})$ which now exhibits two maxima. The maximum at $n_{+} > 0, n_{-} \simeq 0$ 
corresponds to the situation where the wall is driven forward almost exclusively by the 
growing MTs whereas that at $n_{-} > 0, n_{+} \simeq 0$. 

The physical origin of the bi-directional motion in this case can be understood by critically 
examining the interplay of the time-dependence of $n_{+}(t),n_{-}(t)$ and the 
dependence of $c(F),r(F)$ on $n_{+}(t),n_{-}(t)$. Although initially, $n_{+}(0)=N/2=n_{-}(0)$, 
stochastic nature of catastrophe and rescue causes population imbalance in spite of the 
equal values of the corresponding rate constants. Because of the appearance of $n_{+}(t),n_{-}(t)$ 
in eq.(\ref{eq-ppmFfinal}) the population imbalance grows further. Thus, if $n_{+}(t)$ 
keeps increasing with time $t$, the wall continues to move forward; eventually, if $n_{-}$ 
vanishes, $v_{w}$ would become identical to $v_{g}$. However, by that time, because 
of the non-vanishing $c_{0}$, the polymerizing MTs start suffering catastrophe that, 
in turn, increases population of depolymerizing MTs which has a feedback effect. Once 
the majority of the MTs are in the depolymerizing state the wall reverses its direction 
of motion. The velocity of the wall alternates between positive and negative values, 
corresponding to the alternate segments of the trajectory, that characterizes the  
bi-directional motion of the wall.

\begin{widetext}
\begin{figure}[H]
\begin{center}
\begin{tabular}{cc}
(A1)&(B1)\\
\includegraphics[angle=-90,width=0.45\textwidth]{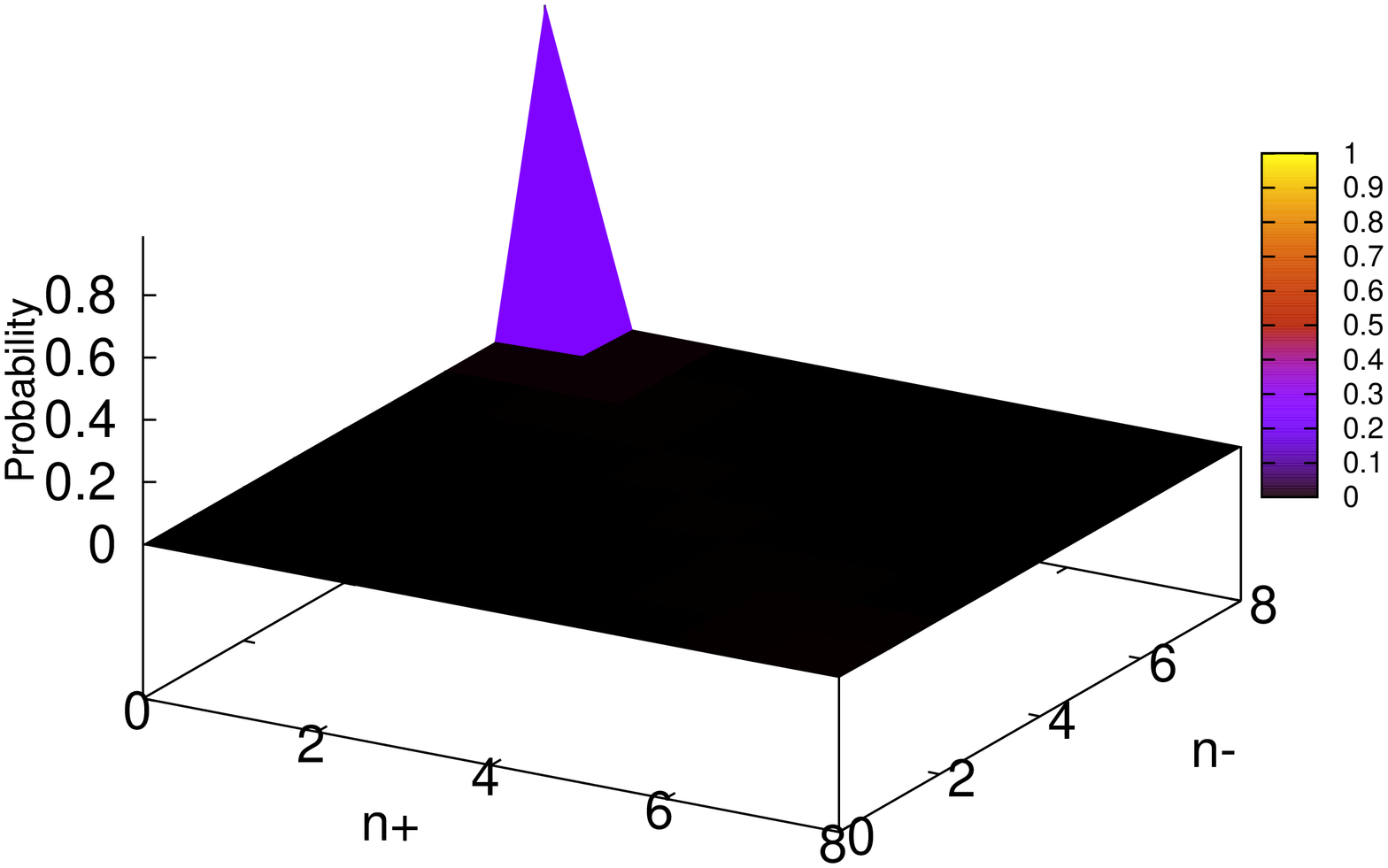}&
\includegraphics[angle=-90,width=0.45\textwidth]{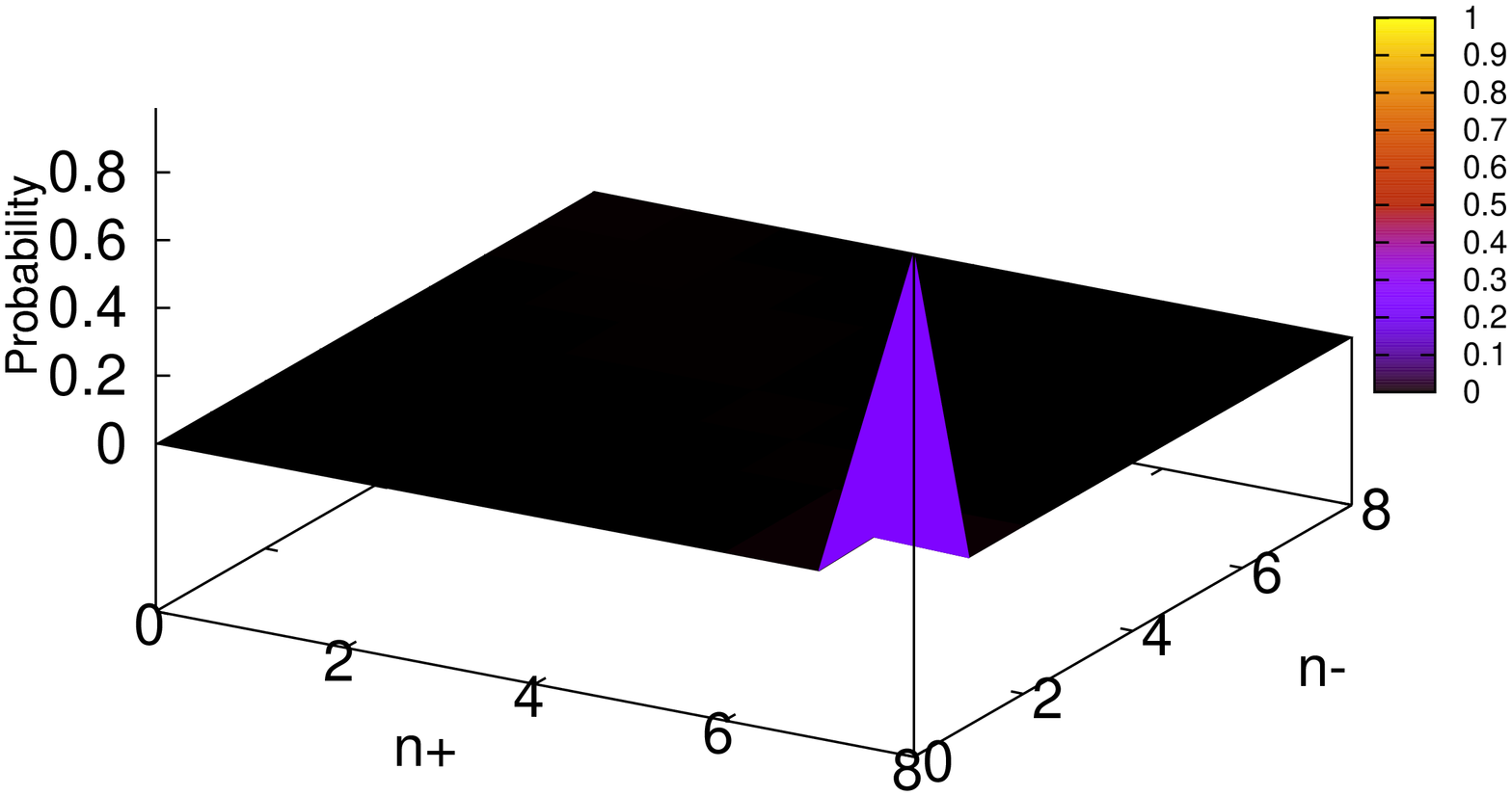}\\
(A2)& (B2)\\
\includegraphics[width=0.45\textwidth]{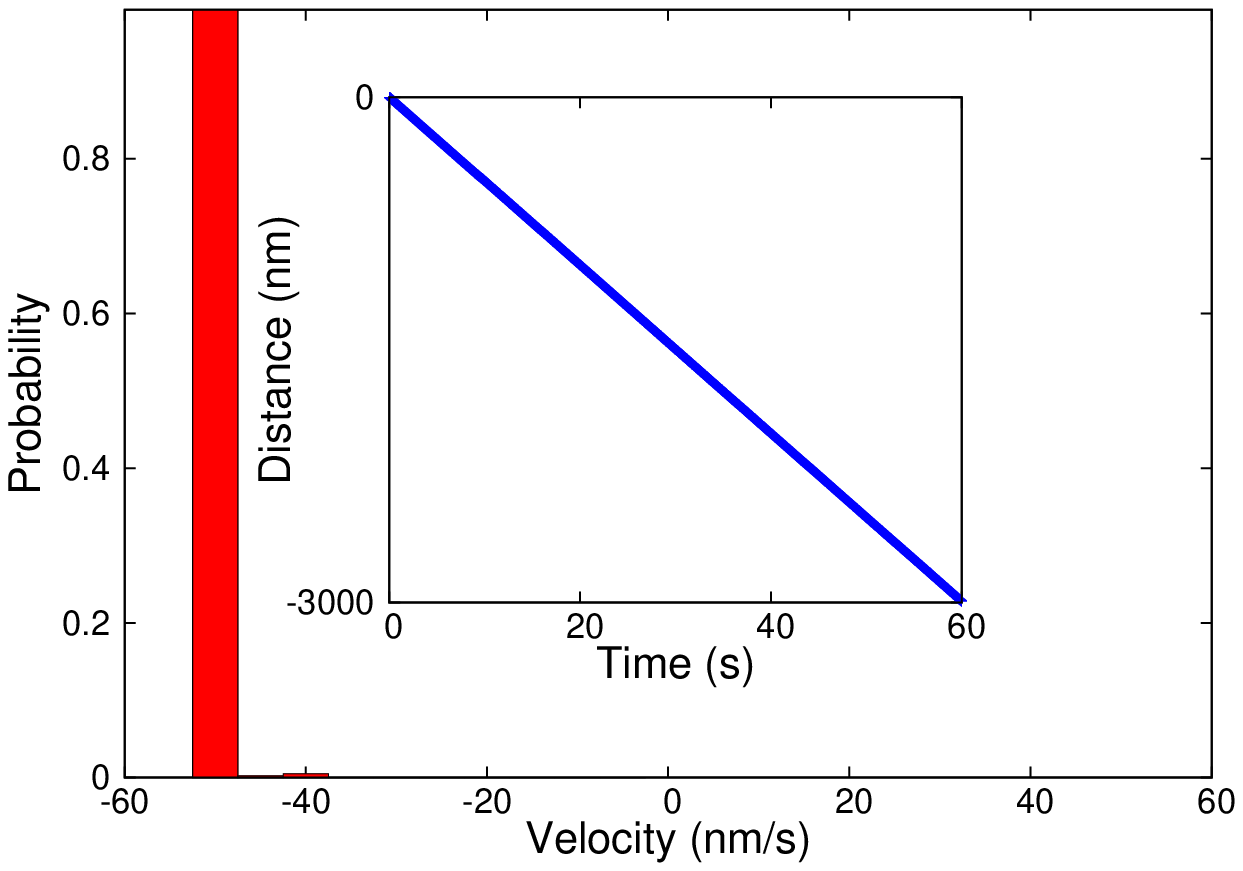}&
\includegraphics[width=0.45\textwidth]{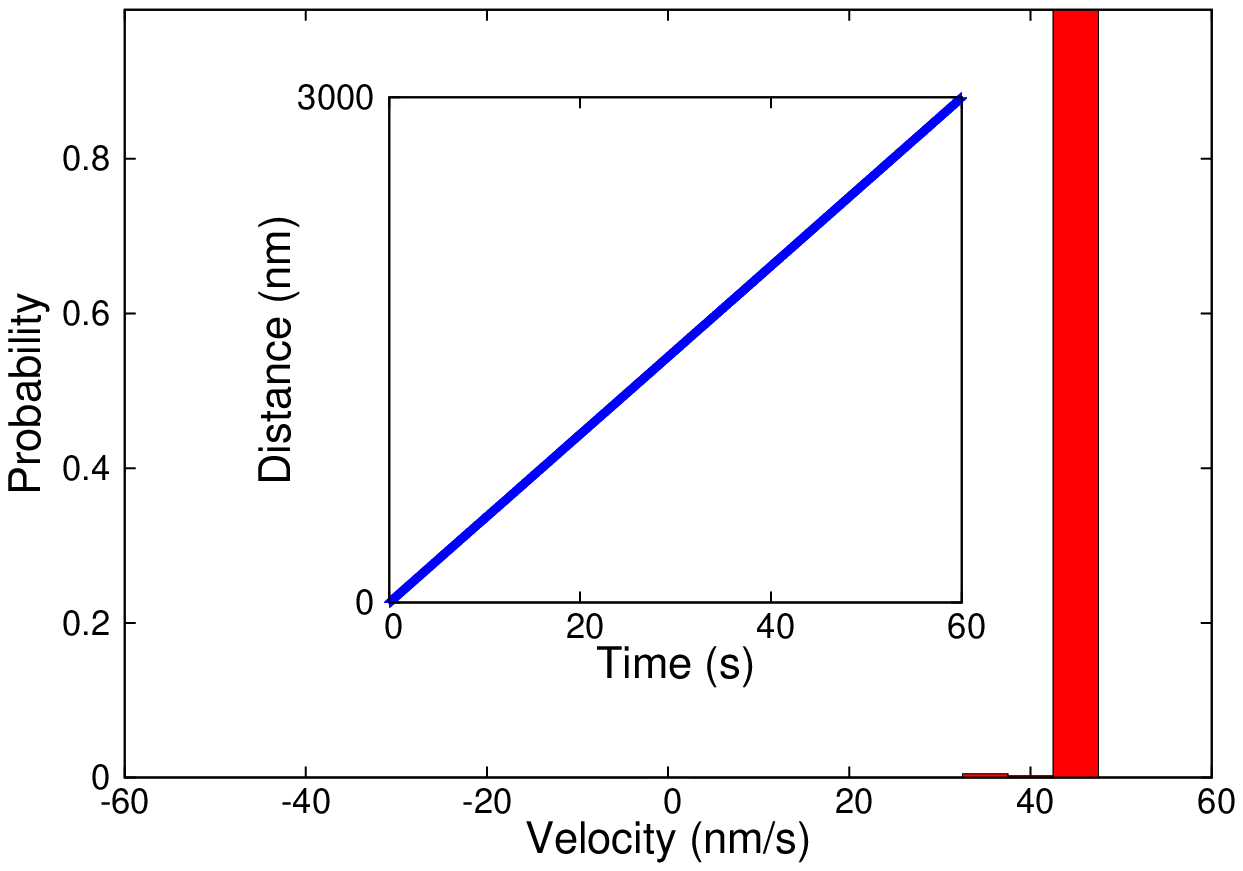}\\
\end{tabular}
\caption{Characteristics of only minus and only plus motion. Numerical value of the parameters, used in the simulation are $ F_{s}=5.0$~pN,$ v_{g}=v_{d}=$50 nm/s,$ c_{0} = r_{0} = 10$ s$^{-1}$. To get only minus or only plus motion we take $ F_{*c} =3.0$~pN , $ F_{*r}=6.0$~pN and vice versa.  
(A1)Probability vs motor number plot has maxima for only shrinking MT number. Corresponding velocity distribution and displacement are shown in (A2) and inset of (A2), respectively. (B1)Probability vs motor number plot has maxima for only growing MT number. Corresponding velocity distribution and displacement are shown in (B2) and inset of (B2), respectively.}
\label{fig_mp}
\end{center}
\end{figure}
\end{widetext}

$\bullet$ {\bf Tug-of-war in the `No Motion (0)' state}

Next we chose the parameter values $ F_{*c}=F_{*r} =10.0$~pN and 
$ c_{0} = r_{0} = 2000 s^{-1}$. Note that that catastrophe and 
rescue rates are much higher than those in the previous case. 
Consequently, each MT very frequently switches between the 
polymerizing and depolymerizing states. Moreover, because of the 
much smaller values of $F_{s}$, a MT can exert very small force 
on the wall in both cases. The results of simulation are plotted in 
fig.\ref{fig_m0}. The velocity distribution exhibits a single peak at 
zero velocity which is a signature of the `No Motion (0)' state 
\cite{berger11}. This is a consequence of the tug-of-war between 
the growing and shrinking MTs that is evident in the single maximum 
at $n_{+}=n_{-}$ in the probability distribution $P(n_{+},n_{-})$. 
Tug-of-war does not mean complete stall; small fluctuations around 
the stall position visible in the trajectory gives rise to the non-zero 
width of the velocity distribution around zero velocity. The underlying 
physical processes in this case are almost identical to those in the 
case of bi-directional motion except for crucial difference arising 
from the relatively higher values of $c_{0} = r_{0}$; even before the 
wall can cover a significant distance in a particular direction the 
MTs reverse their velocities. Because of the high frequencies of 
catastrophe and rescue the wall remains practically static except for 
the small fluctuations about this position.

$\bullet$ {\bf Only plus (+) motion and only minus (-) motion}

Finally, for the sake of completeness, we explored asymmetric behavior of the model by choosing  
$F_{*c} < F_{*r} $ in one case and $ F_{*c} > F_{*r} $ in the other so that in one case the wall exhibits  
only minus motion whereas in the other case it exhibits only plus motion. The simulation results 
are plotted side-by-side for these two cases in fig.\ref{fig_mp}. In this case the typical trajectories 
indicate motion either only in the plus direction or only in the minus direction. This observation is 
also consistent with the  single sharp peaks at $V=+ 50$ nm/s and  $V=- 50$ nm/s in the distribution 
of velocities of the wall. The maximum at  $n_{-} > 0, n_{+} \simeq 0$ corresponds to the situation 
where the wall is driven backward almost exclusively by the shrinking MTs whereas in another case  
$n_{+} > 0, n_{-} \simeq 0$  the wall is driven forward by the growing MTs.


$\bullet$ {\bf Phase diagram}

\begin{figure}[H]
\includegraphics[angle=0,width=0.85\columnwidth]{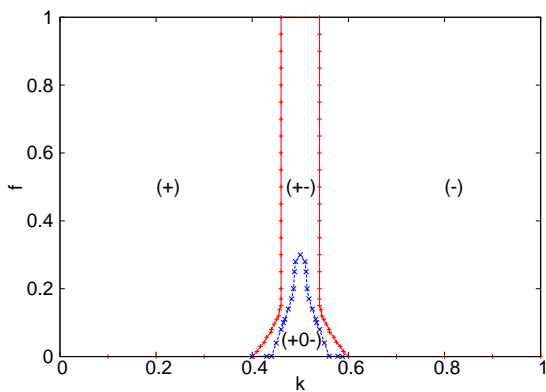}\\
\caption{\bf All the distinct states of motility of the wall 
for the case $F_{*c}=F_{*r}$ are displayed on the $f-k$ phase 
diagram (see text for the definitions of $f$ and $k$). The 
phases are labelled by the corresponding symbols (see text 
for the meaning of these symbols). The re-entrance phenomenon 
exhibited for $f \lesssim 0.23$ is also explained in the text.}
\label{fig-phase}
\end{figure}

The movement of the wall depends crucially on two important 
dimensionless parameters, namely, 
$k=c_{0}/(c_{0}+r_{0})$ and $f=F_{s}/(F_{s}+F_{\star})$. 
For the sake of simplicity and for reducing the number of 
parameters, here we have assumed $F_{*c}=F_{*r}=F_{*}$. 
We have varied these model parameters over wide range of 
values and, for each set of values, we have identified the 
state of movement of the wall. The corresponding observations 
for the symmetric case $F_{*c}=F_{*r}=F_{*}$, are summarized 
in fig.\ref{fig-phase} in the form of a phase diagram on the 
$f-k$ plane. Since no new motility state, other than those 
observed in the symmetric case $F_{*c}=F_{*r}$ are observed 
in the more general asymmetric case $F_{*c} \neq F_{*r}$, we 
have not drawn the higher-dimensional phase diagram. 

For high force ratio ($ f \gtrsim 0.23 $) we observe transition 
from (+) to (-) region through the (+-) region in between, 
as the ratio $k$ is varied.  
In contrast, for $f \lesssim 0.23$ the system exhibits ``re-entrance''; 
as $k$ increases from $0.4$ to $0.6$, first a transition from 
the motility state (+-) to the state (+0-) takes place and, 
at a somewhat higher value of $k$ a re-entrance to the state 
(+-) occurs. This re-entrance phenomenon disappears at $f \gtrsim 0.23$.


\section{Summary and conclusion}

In this paper we have introduced a theoretical model for studying the cargo-mediated 
collective kinetics of polymerization and depolymerization of a bundle of parallel MTs 
that are not bonded laterally with one another.  
Carrying out computer simulations of the model we have identified and characterized  
the motility states of a hard wall-like cargo that is pushed and pulled by this MT bundle 
that is oriented perpendicular to the plane of the wall. Among these motility states, 
one corresponds to ``no motion'' (except for small fluctuations); it arises from the 
``tug-of-war'' between the polymerizing and depolymerizing MTs. The wall exhibits 
a {\it bi-directional} motion in another motility state.  
The qualitative features of the characteristics of this bi-directional 
state of motility of the wall are similar to those observed in 
bi-directional motion of vesicular cargo driven by antagonistic motor 
proteins.
But, as we have argued here, the physical origin 
of the bidirectional motion in these two systems are completely different. In the 
case of motor protein-driven vesicular cargo the bi-directional motion is caused by 
an interplay of the stall force of the motors and the load-dependent detachment 
of the motor from its track \cite{berger11}. In contrast, in our model of MT-driven 
wall, the bi-directional motion arises from a subtle interplay of the stall force of the 
MTs and the load-dependent depolymerization of the MTs. This result should be regarded 
also as a ``proof-of-principle'' that for bi-directional motion of the chromosomal 
cargo active participation motor proteins is not essential; in contrast to past claims 
in the literature (see ref.\cite{sutradhar14} for a very recent claim), the polymerization 
/ depolymerization kinetics of the MTs are adequate to cause bidirectional motion provided 
the load-dependence of their rates are properly taken into account. The latter principle 
will be elaborated further elsewhere \cite{ghanti14b} in the context of chromosome 
segregation with a more detailed theoretical model which also treats the kinetochore wall 
as a soft elastic object \cite{zelinski13}.

\section*{Acknowledgement}
Work of one of the authors (DC) has been supported by Dr. Jag Mohan Garg 
Chair professorship, by J.C. Bose National Fellowship and by a research grant from the 
Department of Biotechnology (India).


 \end{document}